\documentstyle[aps,preprint,epsfig]{revtex}

\def \beq {\begin{equation}}
\def \eeq {\end{equation}}

\begin{document}

\draft

\title{On the singularities of gravity in the presence of 
non-minimally coupled scalar fields}

\author{L.R. Abramo$^1$,
L. Brenig$^2$,
E. Gunzig$^{2,3}$,
Alberto Saa$^{2,4}$ \footnote{e-mails: 
{
abramo@fma.if.usp.br, 
lbrenig@ulb.ac.be, 
egunzig@ulb.ac.be, 
asaa@ime.unicamp.br}}}

\address{1)
Instituto de F\'\i sica, Universidade de S\~ao Paulo,
CP 66318, 05315-970 S\~ao Paulo, SP, Brazil
}

\address{2)
RggR, Universit\'e Libre de Bruxelles, 
CP 231, 1050 Bruxelles, Belgium.
}
\address{3)
Instituts Internationaux de Chimie et de Physique Solvay, 
CP 231, 1050 Bruxelles, Belgium. 
}

\address{4)
IMECC -- UNICAMP,
C.P. 6065, 13081-970 Campinas, SP, Brazil.}

\maketitle

\begin{abstract}
We investigate the robustness of some recent results obtained
for homogeneous and isotropic cosmological models with 
conformally coupled scalar fields.
For this purpose, we investigate anisotropic homogeneous
solutions of the models described by the action
$$
S=\int d^4x \sqrt{-g}\left\{F(\phi)R - \partial_a\phi\partial^a\phi
-2V(\phi) \right\},
$$
with general $F(\phi)$ and $V(\phi)$.
We show that such a class of models leads generically to geometrical
singularities if for some value of $\phi$, $F(\phi)=0$,
rendering previous cosmological results obtained
for the conformal coupling case highly unstable.
We show that stable models can be obtained for
suitable choices of $F(\phi)$ and $V(\phi)$.
Implications for
other recent results are also discussed.

\end{abstract}

\section{Introduction}
Recently\cite{PRD,IJTP1}, we have investigated the dynamics of homogeneous and
isotropic solutions of the cosmological model described by the action:
\beq
\label{act}
S=\int d^4x \sqrt{-g}\left\{F(\phi)R - \partial_a\phi\partial^a\phi
-2V(\phi) \right\},
\eeq
with $F(\phi)=1-\frac{1}{6}\phi^2$, corresponding to 
the so-called conformal
coupling, and $V(\phi)=\frac{m}{2}\phi^2-\frac{\Omega}{4}\phi^4$.
Some novel dynamical behaviors were identified: 
superinflation regimes, 
a possible 
avoidance of big-bang singularities through classical birth of the universe 
from empty Minkowski space, spontaneous entry into and exit from inflation,
and a cosmological history suitable 
for describing quintessence. Since one of the proposals of 
inflationary 
models is to describe our universe without finely-tuned parameters,
these results would be viable 
only if they are robust against small
perturbations in initial conditions and in the model itself. This is the
question to be addressed here. 

We study the robustness of our
previous results by taking two generalizations of the model
considered previously: we relax the isotropy
requirements (perturbations in the initial conditions)
and we consider a general coupling $F(\phi)$ (perturbations in the
model parameters). Models with more general $F(\phi)$ have
been considered recently\cite{G}. Our results
show that the model is not robust.
Its main properties are radically changed, even for small disturbances
in initial conditions and in the model itself, 
due to the appearance of real, gravitational singularities
that are dynamically unavoidable in general. The singularities
 are, essentially, of two types.
The first one corresponds to the hypersurfaces $F(\phi)=0$. 
It is not present in the
isotropic case, and it implies that all previous homogeneous
and isotropic solutions
passing from the $F(\phi)>0$ to the  $F(\phi)<0$ region are
extremely unstable against anisotropic perturbations.
The second type of singularity corresponds to $F_1(\phi)=0$, with
\beq
\label{f1}
F_1(\phi) = F(\phi)+\frac{3}{2}\left(F'(\phi)\right)^2,
\eeq
and it is present even for the homogeneous and isotropic cases. 
Although for small deviations
of the conformal coupling the latter singularities are 
typically very far from the region of
interest, in the general case they can alter qualitatively the global 
dynamics of the model due to restrictions that it imposes on the
phase space. Again, the persistence of some of our previously described
results, in particular
the ones concerning
heteroclinic and homoclinic solutions, are challenged.

Both kinds of singularities have already been described 
before. To the best of our knowledge, Starobinski\cite{Starobinski} was the
first to identify the singularity corresponding to the hypersurfaces
$F(\phi)=0$, for the case of conformally coupled anisotropic solutions.
Futamase and co-workers\cite{Futamase} identified both
singularities
in the context of chaotic inflation in $F(\phi)=1-\xi\phi^2$ theories
(See also \cite{s2}).
The first singularity is always present for $\xi>0$ and the second one for 
$0<\xi<1/6$.
Our conclusions are, however, more general since we treat the case of
general $F(\phi)$ and our results are based on the analysis of
true geometrical invariants. Our main result is that the
system governed by (\ref{act}) is {\em generically} singular on both
hypersurfaces $F(\phi)=0$ and $F_1(\phi)=0$. Here, {\em generically}
means that it is possible to construct non-singular models if
one fine-tunes $F(\phi)$ and $V(\phi)$, as we will show below.

One can advance that there are some geometrically 
special regions on the phase space of the
model in question by a very 
simple analysis of the equations derived from the 
action (\ref{act}).
They are the Klein-Gordon equation
\beq
\label{kg}
\Box\phi - V'(\phi) +\frac{1}{2}F'(\phi)R=0,
\eeq
and the Einstein equations
\begin{eqnarray}
\label{ee}
F(\phi)G_{ab} &=& (1+F''(\phi))\partial_a\phi\partial_b\phi \nonumber \\ &-& 
\frac{1}{2}g_{ab}\left[ (1+2F''(\phi))\partial_c\phi\partial^c\phi 
+2V(\phi)\right] - F'(\phi)\left(g_{ab}\Box\phi - \nabla_a\phi\nabla_b\phi 
\right).
\end{eqnarray}
We will consider here the simplest anisotropic homogeneous cosmological
model, the Bianchi type I, whose spatially flat metric is given by
\beq
\label{metric}
ds^2 = -dt^2 + a^2(t)dx^2 + b^2(t)dy^2 + c^2(t)dz^2. 
\eeq 
The dynamically relevant quantities here are
\beq
H_1 = \frac{\dot{a}}{a}, \quad H_2 = \frac{\dot{b}}{b}, \quad
{\rm and\ } H_3 = \frac{\dot{c}}{c}\ .
\eeq
For such a metric and a homogeneous scalar field $\phi=\phi(t)$, after
using the Klein-Gordon Eq. (\ref{kg}), 
Eq. (\ref{ee}) can be written as
\begin{eqnarray}
\label{ec}
F(\phi)G_{00} &=& \frac{1}{2}\dot{\phi}^2 + V(\phi) - F'(\phi)\left( 
H_1+H_2+H_3\right)\dot{\phi}, \\
\label{e1}
\frac{1}{a^2}F(\phi)G_{11} &=& \frac{1+2F''(\phi)}{2}\dot{\phi}^2 -
V(\phi) - F'(\phi)\left( H_1\dot{\phi} + V'(\phi) -\frac{F'(\phi)}{2}R\right),
\\\label{e2}
\frac{1}{b^2}F(\phi)G_{22} &=& \frac{1+2F''(\phi)}{2}\dot{\phi}^2 -
V(\phi) - F'(\phi)\left( H_2\dot{\phi} + V'(\phi) - \frac{F'(\phi)}{2}R\right),
\\\label{e3}
\frac{1}{c^2}F(\phi)G_{33} &=& \frac{1+2F''(\phi)}{2}\dot{\phi}^2 -
V(\phi) - F'(\phi)\left( H_3\dot{\phi} + V'(\phi) - \frac{F'(\phi)}{2}R\right).
\end{eqnarray}
It is quite simple to show that Eqs. (\ref{e1})-(\ref{e3}) are
not compatible, in general,
 on the hypersurface $F(\phi)=0$.  Subtracting
(\ref{e2}) and (\ref{e3}) from (\ref{e1}) we have, on 
such hypersurface, respectively,
\beq
F'(\phi)(H_1-H_2)\dot{\phi} = 0,\ {\rm and\quad }
F'(\phi)(H_1-H_3)\dot{\phi} = 0.
\eeq
Hence, 
they cannot be fulfilled in general for anisotropic metrics. As it
will be shown, it indeed corresponds to an unmovable (in the Painlev\'e
sense\cite{Painleve}) geometrical 
singularity which cannot be prevented in general 
by requiring that $F'(\phi)=0$ 
or $\dot{\phi}=0$ on the hypersurface.

As to the second singularity we have, after taking the trace 
of the Einstein
equations, that:
\beq
\label{r}
R = R(\phi,\dot{\phi}) = \frac{1}{F_1(\phi)}\left(4V(\phi) +
3V'(\phi)F'(\phi) - (1+F''(\phi))\dot{\phi}^2 \right).
\eeq
Inserting Eq. (\ref{r}) in the Klein-Gordon Eq. (\ref{kg}), one
can see that it contains terms which are singular for $F_1(\phi)=0$. Again, 
as we will see, this corresponds
to an  unmovable geometrical 
singularity, and it cannot be eliminated, in general,
 by demanding that $F'(\phi)=0$ 
on the hypersurface $F_1(\phi)=0$. In both the  
hypersurfaces $F(\phi)=0$ and
$F_1(\phi)=0$ the Cauchy problem is ill-posed, since one cannot
choose general initial conditions.

The hypersurfaces $F(\phi)=0$ and
$F_1(\phi)=0$ also prevent the global definition of an Einstein frame
for the action (\ref{act}), defined by the transformations
\begin{eqnarray}
\label{eg1}
\tilde{g}_{ab} &=& F(\phi)g_{ab}, \\
\label{eg2}
\left(\frac{d\tilde{\phi}}{d\phi}\right)^2 &=& \frac{F_1(\phi)}{2F(\phi)^2}.
\end{eqnarray}
It is well known that in the Einstein frame the Cauchy problem is well posed. Again,
the impossibility of defining a global Einstein frame shed some doubts
about the general Cauchy problem.
Moreover, the standard perturbation theory for helicity-2 and helicity-0
excitations,
derived directly from Eqs. (\ref{eg1})-(\ref{eg2}), 
fails in both hypersurfaces\cite{G}.

\section{The singularities}

In order to check the geometrical nature of these singular 
hypersurfaces, let
us consider the Einstein Eqs. (\ref{ec})-(\ref{e3}) in
detail. For the metric (\ref{metric}), we have the following
identities
\begin{eqnarray}
\label{gg}
G_{00} &=& H_1H_2 + H_2H_3 + H_1H_3, \nonumber \\
G_{11} &=& a^2\left( \dot{H}_1 + H_1(H_1+H_2+H_3) - \frac{1}{2}R\right),  
\nonumber\\
G_{22} &=& b^2\left( \dot{H}_2 + H_2(H_1+H_2+H_3) - \frac{1}{2}R\right), \\
G_{33} &=& c^2\left( \dot{H}_3 + H_3(H_1+H_2+H_3) - \frac{1}{2}R\right),  
\nonumber\\
R &=& 2\left( \dot{H}_1 + \dot{H}_2  + \dot{H}_3  
+ {H}_1^2 + {H}_2^2 + {H}_3^2 + 
 H_1H_2 + H_2H_3 + H_1H_3
\right).  \nonumber
\end{eqnarray}
After using expressions (\ref{gg}) and introducing the
new dynamical variables $p=H_1+H_2+H_3$, $q=H_1-H_2$, and 
$r=H_1-H_3$, Einstein Eqs. can be cast in the form
\begin{eqnarray}
\label{ec1}
E(\phi,\dot{\phi},p,q,r)&=&-\frac{1}{3}F(\phi)\left( 
p^2 + qr - q^2 - r^2
\right) + \frac{\dot{\phi}^2}{2} + V(\phi) - pF'(\phi)\dot{\phi} = 0,\\
\label{eq}
\dot{q} &=&  - \left(p+\frac{F'(\phi)}{F(\phi)}\dot{\phi}\right)q, \\
\label{er}
\dot{r} &=&  - \left(p+\frac{F'(\phi)}{F(\phi)}\dot{\phi}\right)r, \\
\label{ep}
-2F_1(\phi)\dot{p} &=& (F(\phi)+2F'(\phi)^2)p^2+
\frac{3}{2}(1+2F''(\phi))\dot{\phi}^2 - 3V(\phi) - 3 F'(\phi)V'(\phi) 
\nonumber \\
 & & - p\dot{\phi}F'(\phi) 
  + (F(\phi)+F'(\phi)^2)(q^2 + p^2 - qr)
\end{eqnarray}
Using the energy constraint (\ref{ec1}), this last equation can
be put in the form
\begin{eqnarray}
\label{ep1}
\dot{p} &=& - \left(p+\frac{F'(\phi)}{F(\phi)}\dot{\phi}\right)p + 
3 \frac{ F(\phi) +\frac{1}{2}(F'(\phi))^2}{F(\phi)F_1(\phi)}
V(\phi) \nonumber \\
&& -\frac{3}{4}\left(
\frac{2F(\phi)F''(\phi)-(F'(\phi))^2}{F(\phi)F_1(\phi)}\right)\dot{\phi}^2
+\frac{3}{2}\frac{F'(\phi)}{F_1(\phi)}V'(\phi).
\end{eqnarray}
The 
Klein-Gordon equation (\ref{kg}) reads simply
\beq
\label{kg1}
\ddot{\phi} + p\dot{\phi} + V'(\phi) - \frac{F'(\phi)}{2}R(\phi,\dot{\phi})=0,
\eeq
with $R(\phi,\dot{\phi})$ given by Eq. (\ref{r}).
The energy constraint, equation (\ref{ec1}), is evidently compatible with
the other ones. Indeed $E(\phi,\dot{\phi},p,q,r)=0$ is an invariant
surface since one has that
\beq
\frac{d}{dt}E(\phi,\dot{\phi},p,q,r) = -\left(
2p + \frac{F'(\phi)}{F(\phi)}\dot{\phi}
\right)E(\phi,\dot{\phi},p,q,r)
\eeq
along solutions of Eqs. (\ref{eq}), (\ref{er}), (\ref{ep1}), and (\ref{kg1}).
Note that Eqs. (\ref{ep1}) and (\ref{kg1}) are decoupled
from the equations for $\dot{q}$ and $\dot{r}$. Equations (\ref{eq}) and
(\ref{er}) are, hence, linear first order equations, and they could
be easily integrated after the solutions of (\ref{ep1}) and (\ref{kg1})
have been found.
Moreover, since one has $r\dot{q}-q\dot{r}=0$, $q(t)/r(t)$ is a constant
of motion
fixed only by the initial conditions.
Suppose the initial ratio is $q(0)/r(0)=\gamma$: this would imply that
$(H_1-H_2)=\gamma(H_1-H_3)$ for all $t$, leading to, for instance,
$c^\gamma(t) \propto a^{\gamma-1}(t)b(t)$ in the
metric (\ref{metric}). This simplification
is a consequence of the  scalar character of our source field,
and it does not suppose any
loss of generality in our arguments.

A closer analysis of Eqs. (\ref{eq})-(\ref{ep1}) reveals the presence of the
singularities. In general, the right-hand side of these equations
diverge for $F(\phi)=0$ and for $F_1(\phi)=0$. One can check that
these divergences are indeed related to real geometrical singularities
by considering the Kretschman invariant $I=R_{abcd}R^{abcd}$, which
for the metric (\ref{metric}) is given by
\beq
I = 4\left(
\left(\dot{H}_1+H_1^2\right)^2 + 
\left(\dot{H}_2+H_2^2\right)^2 + 
\left(\dot{H}_3+H_3^2\right)^2 
+ 
H_1^2H_2^2 + H_1^2H_3^2 + H_2^2H_3^2
\right).
\eeq
As one can see, $I$ is the sum of non negative terms.
Moreover, any divergence of the variables $H_1$, $H_2$, $H_3$, or of
their time derivatives, would suppose a divergence in $I$,
characterizing a real geometrical singularity. Since 
the relation between the variables $p$, $q$, $r$, and
$H_1$, $H_2$, $H_3$ is linear, any divergence of the
first, or of their time derivative, will suppose a divergence
in $I$. 

Suppose, now, that $F(\phi_{0})=0$, and that
$F(\phi)$ is (real) analytical for $\phi=\phi_{0}$. In this case,
$F'(\phi)/F(\phi)$ diverges as $(\phi-\phi_{0})^{-1}$ 
near $\phi_{0}$ for nonvanishing $q$ and $r$, 
rendering $\dot{q}$ and $\dot{r}$ 
divergent by (\ref{eq}) and (\ref{er}), unless $p$ also
diverges in order to keep $\left(p+\frac{F'(\phi)}{F(\phi)}\dot{\phi}\right)$ 
finite on $\phi_{0}$. In both cases, $I$ diverges.
There is no dynamical restriction to ensure that $\dot{\phi}$
vanishes on $\phi_{0}$ - it can take any value compatible
with the energy constraint (\ref{ec1}). Indeed, the latter implies that
on $\phi_{0}$
\beq
\frac{\dot{\phi}^2}{2} - p F'(\phi_{\rm 0})\dot{\phi} + V(\phi_{\rm 0}) = 0.
\eeq
There is no way of having $\dot{\phi}=0$ on $\phi_0$, unless
$V(\phi_0)=0$, and even in this case, 
$\dot{\phi} =p F'(\phi_0)$ is also possible.
Note that the hypothesis of $F(\phi)$ analytical at
$\phi_0$ is not a necessary one. For any differentiable function 
$F(\phi)$ with a zero in $\phi_0 $ one has 
$|F(\phi)| = |\int_{\phi_0}^\phi F'(s) ds| \le k|\phi-\phi_0|$, with
$k=\max_{s\in[\phi_0,\phi]}|F'(s)|$, implying that
$|F'(\phi)/F(\phi)| \ge |\phi-\phi_0|^{-1}
|F'(\phi)|/k$.
Since $F'(\phi)$ is assumed to be continuous, the last ratio tends to
1 when $\phi\rightarrow\phi_0$, implying the divergence 
of  $F'(\phi)/F(\phi)$ in that limit.

Now, let us suppose  $F_1(\phi_{1})=0$. If $F'(\phi_1)\ne 0$, the
right-hand side of Eq. (\ref{ep1}) diverges. The
vanishing of $F'(\phi_1)$ implies, by (\ref{f1}), that
$F(\phi_1)=0$, and the arguments of the last paragraph can be repeated.

A singularity-free model can be constructed by demanding that
$F(\phi_0) =F'(\phi_0) = 0$, by choosing a $V(\phi)$ that goes to 0 
at a proper rate when $\phi\rightarrow\phi_0$,  
and by demanding that
$F_1(\phi)$ have no other zeros than the ones of  $F(\phi)$.
Models for which $F(\phi) = \zeta\phi^{2n}$ and
$V(\phi) = \alpha\phi^{2(2n-1)} + {\rm\ high\ order\ terms}$, 
for instance, fulfill these
requirements. However,
such a highly fine-tuned class of model is of no physical interest here, 
since it does not contain $F(\phi)>0$ and $F(\phi)<0$ regions and consequently
has no solution for which the effective gravitational constant
$G_{\rm eff}$ changes it sign along the
cosmological history. The stability of such solutions were the
starting point of the analyses of the pioneering work \cite{Starobinski}
and of the present one as well. 
Note that by Eq. (\ref{f1}), models with an $F(\phi)<0$ region will 
allways have
 singularities of the type $F_1(\phi)=0$. This fact shall be
taken into account to better understand the recently proposed
dynamical stability of the $F(\phi)<0$ region\cite{stability}.

\section{Conclusion}

The singularities described in the precedent section imply that
the model presented in \cite{PRD,IJTP1} is not robust, since 
our main conclusions were a 
consequence of very especial initial conditions.
For instance, all homogeneous and isotropic solutions crossing the
$F(\phi)=0$ hypersurface are extremely unstable against anisotropic
perturbations. By Eqs. (\ref{eq}) and (\ref{er}), any deviation 
from perfect  isotropy
(expressed by nonvanishing $q$ and $r$ variables) for these solutions, 
however small, will lead catastrophically to a
geometrical singularity. Many of the novel dynamical behaviors
presented in \cite{PRD,IJTP1} depend on these solutions. This is 
the case, for instance, of some solutions exhibiting 
superinflation regimes. The heteroclinic and homoclinic
solutions identified in \cite{PRD,IJTP1} can cross the 
$F(\phi)=0$ hypersurface and, hence,
they also suffer the same
instability against anisotropic perturbations. The homoclinic
solutions were considered as candidates to describe a non-singular 
cosmological history, with the big-bang singularity being
avoided through a classical birth of the universe from empty
Minkowski space. Apart from $F(\phi)=0$ singularities, 
these solutions are also affected by the singularities of type 
$F_1(\phi)=0$.
Suppose that the conformal coupling is disturbed by a very
small negative term: $F(\phi)=1-(\frac{1}{6}-\epsilon)\phi^2$.
The $F_1(\phi)=0$ singularities will be near the 
$\phi=\pm 1/\sqrt{\epsilon}$ hypersurfaces. Although they are
located far from the $F(\phi)=0$ regions, they alter the
global structure of the phase-space. In this case, they restrict
the existence of homoclinics, rendering a non-singular 
cosmological history more improbable.

The singularities do not affect the conclusions obtained
by considering solutions inside the $F(\phi)>0$ region.
The asymptotic solutions presented in \cite{IJTP1}, for
instance, are still valid. The conclusion that
for large $t$ the dynamics of any solution (inside $F(\phi)>0$)
tends to an infinite diluted matter dominated universe 
remains valid. Moreover, for small anisotropic deviations
($q$ and $r$ small in comparision with $p$), Eqs. (\ref{eq}) and (\ref{er})
allow us to conclude that solutions inside $F(\phi)>0$, 
for large $t$, approach exponentially isotropic matter-dominated universe.

\acknowledgements

The authors
wish to thank G. Esposito-Farese for previous discussions concerning the
Cauchy problem for models with $F(\phi)=0$, and
Profs. Albert and Anny Sanfeld for the warm hospitality
in Mallemort, France, where this work was initiated.
They also acknowledge the financial support from the EEC 
(project HPHA-CT-2000-00015), from OLAM - Fondation pour la Recherche
Fondamentale (Belgium), 
from Fondation Science et Environnement (France),
and from CNPq (Brazil).

\end{document}